\documentclass[11pt,preprint]{aastex}

\newcommand{\msun}{M_\odot}

\newcommand{\mbh}{\ensuremath{M_{\rm BH}}}

\newcommand{\lbol}{\ensuremath{L_{Bol}}}

\newcommand{\lledd}{\ensuremath{L_{Bol}/L_{Edd}}}

\newcommand{\dv}{{\ensuremath{\Delta v}}}

\newcommand{\lya}{\ensuremath{{\rm Ly}{\alpha}}}
\newcommand{\lyal}{Ly{\sc $\alpha$}\,$\lambda$1216}

\newcommand{\nv}{N\,{\sc v}}
\newcommand{\nvl}{N\,{\sc v}\,$\lambda$1240}

\newcommand{\ha}{\ensuremath{{\rm H}{\alpha}}}
\newcommand{\hb}{\ensuremath{{\rm H}{\beta}}}

\newcommand{\siiv}{Si\,{\sc iv}}

\newcommand{\oiv}{O\,{\sc iv}]}

\newcommand{\siivoivl}{\siiv+\oiv\,$\lambda$1400}
\newcommand{\civ}{C\,{\sc iv}}
\newcommand{\civl}{C\,{\sc iv}\,$\lambda$1549}

\newcommand{\siiii}{Si\,{\sc iii}]}
\newcommand{\siiiil}{Si\,{\sc iii}]\,$\lambda$1892}
\newcommand{\ciii}{{C\,{\sc iii}]}}
\newcommand{\ciiil}{C\,{\sc iii}]\,$\lambda$1909}

\newcommand{\hei}{He\,{\sc i}}
\newcommand{\heil}{He\,{\sc i}\,$\lambda$5876}
\newcommand{\heii}{He\,{\sc ii}}

\newcommand{\mgii}{Mg\,{\sc ii}}
\newcommand{\mgiil}{Mg\,{\sc ii}\,$\lambda$2798}
\newcommand{\nai}{Na\,{\sc i}}
\newcommand{\nail}{Na\,{\sc i}\,$\lambda\lambda$5890,5896}

\newcommand{\oiii}{[O\,{\sc iii}]}
\newcommand{\oiiil}{[O\,{\sc iii}]\,$\lambda$5007}

\newcommand{\fe}{Fe}
\newcommand{\feii}{Fe\,{\sc ii}}

\newcommand{\hst}{{\it HST}}

\newcommand{\kms}{\ensuremath{{\rm km~s}^{-1}}}

\newcommand{\ergs}{\ensuremath{\mbox{erg s}^{-1}}}
\newcommand{\flux}{erg s$^{-1}$ cm$^{-2}$ }
\newcommand{\fluxl}{erg s$^{-1}$ cm$^{-2}$ \AA$^{-1}$}

\shorttitle{UV-optical Emission-Line Properties}
\shortauthors{Tang et al.}

\begin{document}
\title{The Optical and Ultraviolet Emission-Line Properties of Bright Quasars with Detailed Spectral Energy Distributions}
\author{Baitian Tang,\altaffilmark{1,2,3}
Zhaohui Shang,\altaffilmark{1,4}
Qiusheng Gu,\altaffilmark{2}
Michael S. Brotherton,\altaffilmark{4}
Jessie C. Runnoe,\altaffilmark{4}
}

\altaffiltext{1}{Department of Physics, Tianjin Normal University,
Tianjin 300387, P. R. China; {\it zshang@gmail.com}}
\altaffiltext{2}{Department of Astronomy, Nanjing University,
Nanjing 210093, P. R. China; Key Laboratory of Astronomy and
Astrophysics}
\altaffiltext{3}{ Department of Physics \& Astronomy,
Washington State University, Pullman, WA 99163}
\altaffiltext{4}{Department of Physics and Astronomy, University of
Wyoming, Laramie, WY 82071, USA 
}

\begin{abstract}  

We present measurements and statistical properties of the optical and
ultraviolet emission lines present in the spectra of 85 bright quasars 
which have detailed spectral energy distributions. 
This heterogeneous sample has redshifts up to $z=1.5$\ and
is comprised of three subsamples
that may be of particular utility: ultraviolet excess Palomar-Green
quasars, quasars with far-ultraviolet coverage from FUSE, and
radio-loud quasars selected to have similar extended radio luminosity
originally selected for orientation studies.  Most of the objects have
quasi-simultaneous optical-ultraviolet spectra, with significant
coverage in the radio-to-X-ray wavebands.  The parameters of
all strong emission lines are measured by detailed spectral fitting.
Many significant correlations previously found among quasar emission-line
properties are also present in this sample, e.g., the Baldwin effect,
the optical correlations collectively known as eigenvector 1, and others.
Finally, we use our measurements plus scaling relationships to estimate
black hole masses and Eddington fractions.  We show the mass estimates
from different emission lines are usually in agreement within a
factor of 2, but nearly a third show larger differences.  We suggest using
multiple mass scaling relationships to estimate black hole masses when 
possible, and adopting a median of the estimates as the black hole mass for
individual objects.  Line
measurements and derived AGN properties will be used for future
studies examining the relationships among quasar emission lines and
their spectral energy distributions.

\end{abstract}


%
\section{Introduction}
\label{sec:intro}

Quasars are high-luminosity active galactic nuclei (AGNs), likely 
powered by accretion onto supermassive black holes.  Through a variety of
physical processes, quasars emit continuous radiation from the radio
to $\gamma$-rays.  The primary energy output is the ``Big Blue Bump''
often attributed to emission from a central accretion disk.  
Disk photons are reprocessed
by other surrounding gas, through heating of dust in an obscuring torus
as well as ionizing the broad line region (BLR) and narrow 
line region (NLR).

Emission lines from the BLR and NLR help cool gas ionized by the quasar,
and can be diagnostic of the metallicity, ionizing continuum, the 
local gravitational field, and other fundamental AGN properties.  
While the optical-ultraviolet (UV) spectra of quasars generally resemble each 
other, they show variations in which systematic correlations exist 
among emission-line and continuum properties.


Some emission-line properties depend strongly on continuum luminosity,
such as the emission-line equivalent width (EW), particularly for 
\civ\ \citep{Baldw77}.  This so-called Baldwin effect shows that 
as luminosity increases, the EW decreases.
With a sample of 744 type 1 AGN spanning 6 orders of magnitude in
continuum luminosity, \citet{Dietr02} observed an anti-correlation
between the slope of the Baldwin effect and the ionization energy
of the emission line ion
\citep[see also][]{EspAnd99}.  \citet{BasLao04} found the
correlation between $\rm EW$ and Eddington ratio is even stronger than
the Baldwin effect, and suggested that relationship may be primary.

The strongest trends among emission lines involve a suite of
correlations collectively known as ``eigenvector 1.''
\citet[][hereafter BG92]{BG92}
employed principal component analysis (PCA) to
investigate 87 low-redshift Palomar-Green quasar spectra covering
$\lambda$4300-5700\AA. PCA identifies orthogonal eigenvectors, which are
linear combinations of input observables that correlate with each
other, and which optimally account for the input data variance.  
The first eigenvector of BG92 (BGEV1), which accounts for the
most variance in the emission-line properties, is characterized
primarily by the anti-correlation between the strength of \oiii\ and
optical \feii, with other parameters involved, such as $\rm H\beta$ full
width half maximum (FWHM) and asymmetry.  Boroson (2002) suggested
that BGEV1 is driven by Eddington ratio ($\rm L/L_{Edd}$).  Later,
BGEV1 was expanded to include both ultraviolet emission-line
properties and continuum properties in other wavebands
\citep{Sulen00}: 
(1) FWHM of broad $\rm H\beta$, (2) equivalent width 
ratio of optical \feii~to broad $\rm H\beta$, (3) soft X-ray photon
index, (4) \civ ~$\lambda$1549 broad line profile displacement at half
maximum.  It would be misleading to emphasize just a few line
properties as defining BGEV1, when many are correlated (e.g., Si
III]/C III] and others, see Wills et al.~1999), as well as features of
the SED including the radio and X-ray loudness (see Kellerman et al.
1989, BG92, Corbin 1993, etc.)

Several recent papers \citep[e.g.,][]{Hu08, Gaske09, FerBal99, Dong09}
have argued that BGEV1 can be understood in terms of the fractions of
BLR clouds at high column density, which may be infalling and emit the
optical \feii\ lines.  Cloud column density, in addition to the
Eddington fraction, then governs where a quasar sits on the BGEV1
relationships.  Unraveling these complicated relationships
to the point that they are well understood will take more effort.

Despite indications that components of the BLR and NLR may be undergoing
some degree of bulk infall or outflow, 
there is also strong evidence that the motions
may be considered generally Keplerian \citep[e.g.,][]{Peter91,Wande99,Onken04}
and used to estimate black hole masses.
Reverberation mapping of AGNs determines the time lag between continuum
changes and the response by emission lines from the BLR
\citep[e.g.,][]{Kaspi07, Bentz09},
establishing a 
size scale.  Doppler widths of the variable emission-line components 
establish velocities.  Together the size and speed are used to determine
the mass of the central gravitational black hole.  The size of the 
BLR scales with continuum luminosity (e.g., Kaspi et al.~2000), allowing
scaling relationships to be developed relating continuum, emission-line
FWHM, and black hole mass (e.g., Vestergaard et al. 2002).
Again, the emission lines and underlying continuum emission are related
through fundamental quasar properties.

We have recently presented new radio-to-X-ray SEDs of a sample of 85 quasars
\citep{Shang11}.
A primary virtue of these SEDs is the detailed
quasi-simultaneous spectrophotometry of the optical-ultraviolet
region.  In this paper, we present measurements of the emission-line
properties of this data set following the technique of Shang et
al.~(2007).  Sample selection and data reduction are described in
Section 2.  In section 3, we use scaling relationships to
estimate black hole masses as well as Eddington fractions for each quasar.  
We show in Section 4 that our sample displays the
previously discovered relationships described above.  We briefly discuss our
results in Section 5, including future applications for our
measurements.  In this paper, we use a cosmology with H$_{0}$=70 km
s$^{-1}$ Mpc$^{-1}$, $\Omega_{M}$=0.3, and $\Omega_{\Lambda}$=0.7.

\section{Sample, Data, and Measurements}
\label{sec:data}

We use the sample of \citet{Shang11}.  This SED atlas has a total of 85 
objects from three different subsamples which are described briefly below.  

\begin{enumerate}

\item The `PGX' subsample contains 22 of 23 Palomar-Green (PG) quasars
in the complete sample selected by \citet{Laor94,Laor97} to study the
soft-X-ray regime.  This subsample is UV bright and has  $z \le 0.4$.
The optical-UV region is covered by UV spectra from the Faint Object
Spectrograph (FOS) on {\it Hubble Space Telescope} (\hst) and
quasi-simultaneous ground-based optical spectra from McDonald
Observatory. See \citet{Shang03,Shang07} for details of this
subsample.

\item The `FUSE-HST' subsample has 24 objects, 17 of which come from
the \emph{FUSE} AGN program \citep{Kriss00, Shang05}.  This is a heterogeneous,
UV-bright sample with $z<0.5$.  The SEDs for this subsample have
quasi-simultaneous \emph{FUSE} \citep{Moos00}, \emph{HST}, and Kitt
Peak National Observatory (KPNO) observations. \\

\item The `RLQ' subsample includes nearly 50 quasars originally
assembled to study orientation; all members of the sample have similar
extended radio luminosity which is thought to be isotropic
\citep{Wills95}.  
Both lobe and core-dominant quasars are included to
have a wide range in the ratio of core to extended radio emission, an
indicator of the orientation.
The SEDs have quasi-simultaneous \emph{HST} and McDonald or KPNO
observations.  See \citet{Wills95}, \citet{Netze95} and
\citet{Runno12b} for additional details on this subsample. The blazars
originally included in this sample are excluded in the SED atlas
because of their variability due to optical synchrotron emission from
a beamed jet. \\

\end{enumerate}

\begin{figure}
\includegraphics[angle=270,scale=0.3]{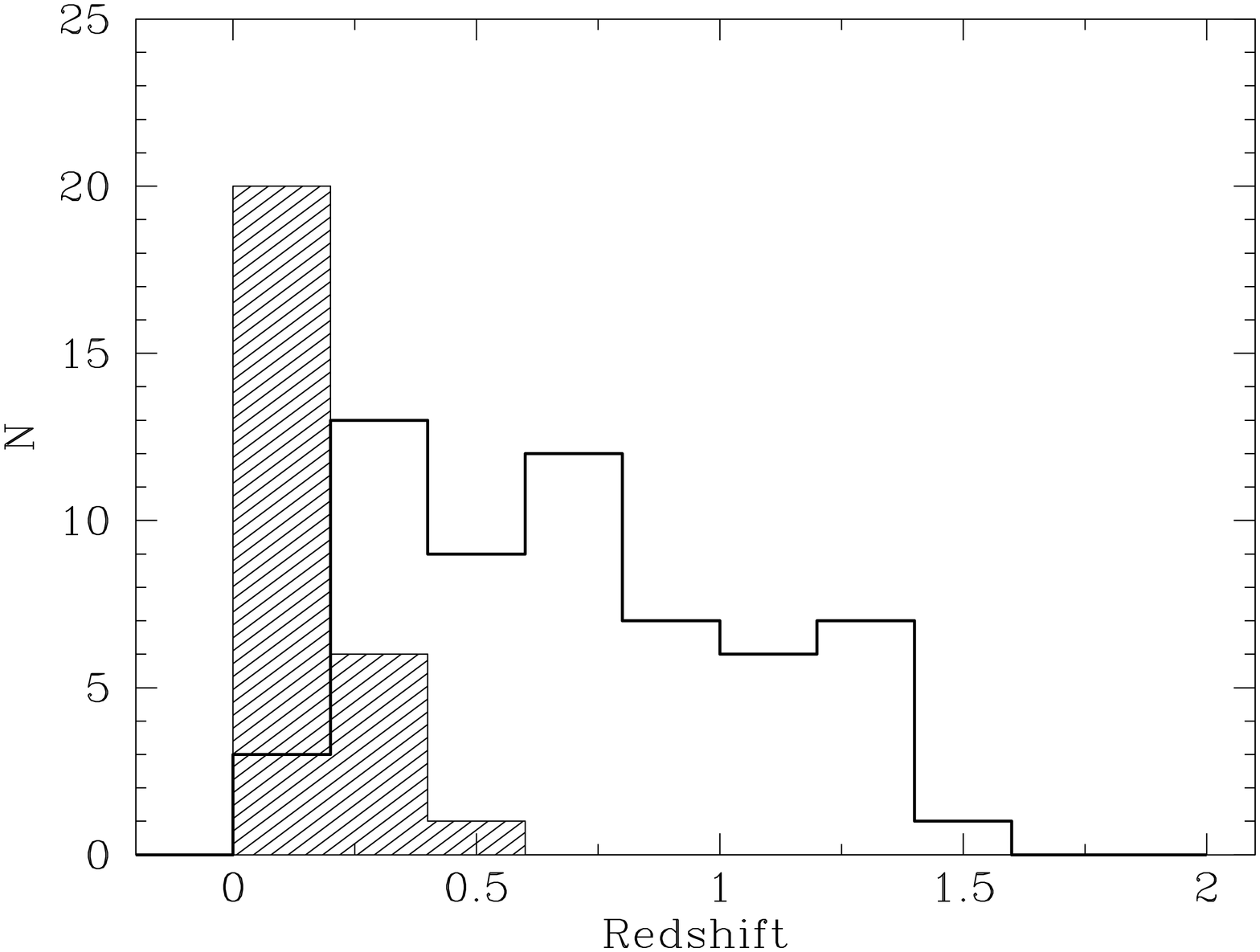}
\caption{Distribution of the sample redshift.  
The shaded area is for radio-quiet
objects, and the thick line is for radio-loud objects.
\label{fg:z}}
\end{figure}

After accounting for duplication of several quasars among the subsamples, 
we total 85 quasars, listed in Table \ref{tb:sample}.  
Figure \ref{fg:z} shows the redshift distribution.

Most of the quasars in our sample have quasi-simultaneous optical and UV
spectra.  All were observed with the \hst\ Faint Object Spectrograph,
primarily the radio-loud and PGX samples, or the Space Telescope
Imaging Spectrograph (STIS), primarily the FUSE sample.  Within a few
weeks of the HST observations, low-resolution ground-based optical
spectrophotometry were obtained at McDonald Observatory or KPNO.  
These quasars are bright in the optical and hence the host galaxy
contamination is negligible as shown by \citet{Shang11}.


The combined ultraviolet-optical spectra were corrected for Galactic
extinction following the
empirical mean extinction law of \citet{Carde89} and the values of
$E(B-V)$ from \citet{Schle98}, assuming $R_V=A_V/E(B-V)=3.1$.  Then
the spectra were shifted to the rest-frame using the redshift of
\oiiil. See \citet{Shang11} for details.

These reddening-calibrated, redshift-calibrated spectra were divided
into several main spectral regions: \lyal, \civl, \ciiil, \mgiil,
\hb~$\lambda$4686, and \ha~$\lambda$6563.  We followed the recipes of
fitting procedures presented by Shang et al.~(2007), although we do
not include the \siivoivl, and \heil\ regions because of low S/N ratio
or missing data. Briefly, we use the IRAF package {\sc specfit} to
find the $\chi^{2}$ minimization between the observed and model
spectra. Each model spectrum consisted of a power-law component and
several Gaussian components. A pseudo-continuum template of
\feii~emission lines was added upon the power-law component in
\mgii~and $\rm H\beta$ regions.  The templates of optical and
ultraviolet \feii~emission lines were derived from the narrow line
Seyfert 1 I Zw1 (BG92; Vestergaard \& Wilkes~2001).  The templates
were allowed to vary in amplitude and velocity width to match the
observed spectra.  Each strong, broad emission line, including
individual lines in doublets \civ\ and \mgii, was fitted with
the combined profile of two Gaussian components, to which no physical
meaning is imbued, but they do reproduce observed line shapes well.  The
parameters of broad emission lines we measured were based on these
Gaussian sums.  

We define the asymmetry parameter following BG92 and
Shang et al.~(2007): \begin{equation} \rm Asymm =
\frac{\lambda_{c}(3/4)-\lambda_{c}(1/4)}{FWHM}.  \end{equation} where
$\rm \lambda_{c}(3/4)$ and $\rm \lambda_{c}(1/4)$ are the wavelength
centers of 3/4 and 1/4 peak flux cuts, respectively.  A positive value
indicates excess light in the blue wing of the line.

\begin{figure}
\epsscale{0.5}
\plotone{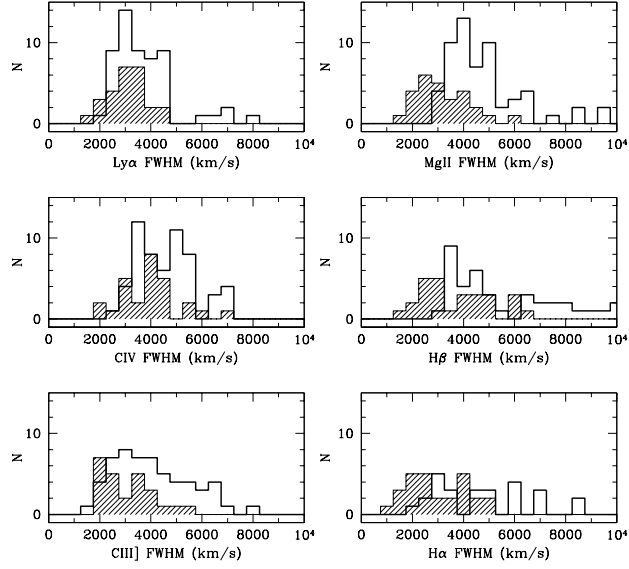}
\caption{Distribution of emission-line FWHM.  The shaded area is for radio-quiet
objects, and the thick line is for radio-loud objects.
\label{fg:fwhmhist}}
\end{figure}

\begin{figure}
\epsscale{0.5}
\plotone{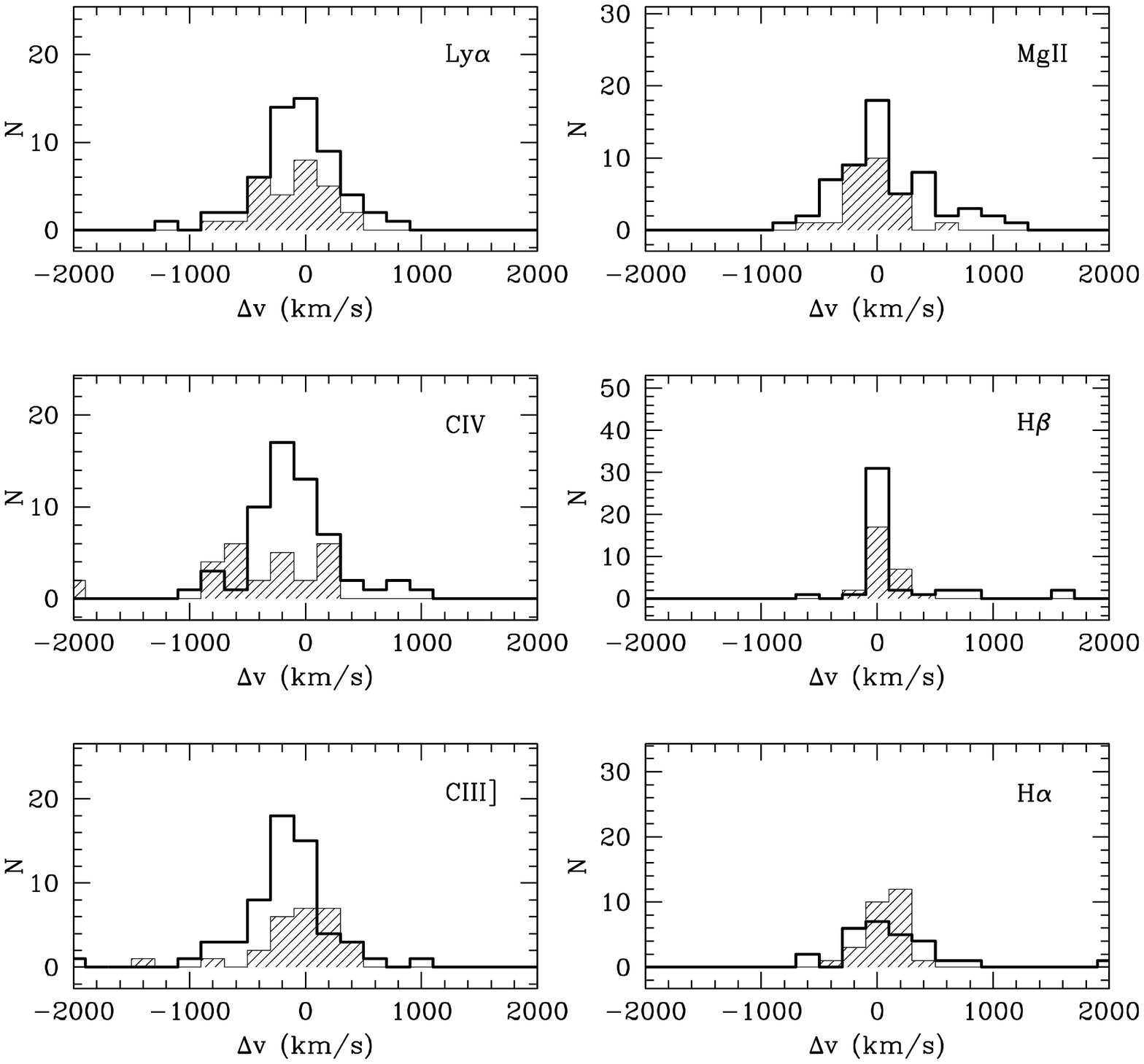}
\caption{Distribution of the emission-line shift relative to \oiiil.
The shaded area is for radio-quiet
objects, and the thick line is for radio-loud objects.
\label{fg:shifthist}}
\end{figure}

\begin{figure}
\epsscale{0.5}
\plotone{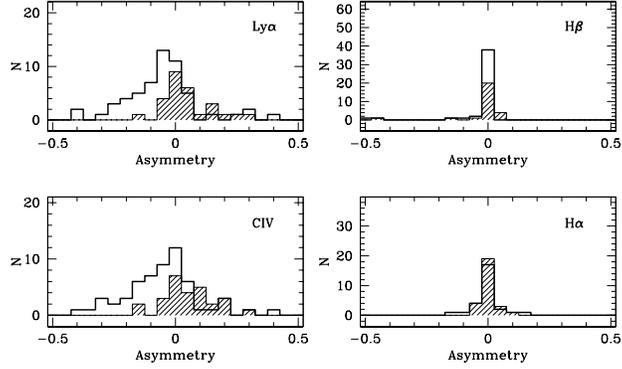}
\caption{
Distribution of emission-line asymmetry.
The shaded area is for radio-quiet
objects, and the thick line is for radio-loud objects.
See \S\ref{sec:data} for definition of the asymmetry
parameter.
\label{fg:asymmhist}}
\end{figure}

We provide our measurements of emission lines in
Table~\ref{tb:stronglinesA}-\ref{tb:stronglinesE}.  The distributions
of some emission-line parameters of this sample are shown in
Figure~\ref{fg:fwhmhist}-\ref{fg:asymmhist}.


\section{\mbh\ and \lledd}
\label{sec:mbh}

Two fundamental parameters can be readily estimated from our
measurements: black hole mass and Eddington fraction.
We use the scaling relationships for single epoch data from
Vestergaard \& Peterson (2006) (for \civ\ and H$\beta$)
and Vestergaard \& Osmer (2009) (for \mgii) 
to estimate black hole masses:

\begin{equation}
\log \mbh(\hb) =\log \left( \left[
\frac{\mbox{FWHM}(\hb)} {1000\,\kms}\right]^2
\left[\frac{\lambda L_{\lambda}(5100 \mbox{\AA})} {10^{44}\,\ergs} 
\right]^{0.50} \right) + (6.91\pm 0.02).
\end{equation}

\begin{equation}
\log \mbh(\mbox{C {\sc IV}}) =\log \left( \left[
\frac{\mbox{FWHM}(\mbox{C {IV}})} {1000\,\kms}\right]^2
\left[\frac{\lambda L_{\lambda}(1350 \mbox{\AA})} {10^{44}\,\ergs} 
\right]^{0.53} \right) + (6.66\pm 0.01).
\end{equation}

\begin{equation}
\mbh(\mbox{Mg {\sc II}}) = 10^{6.86\pm0.55} \left[
\frac{\mbox{FWHM}(\mbox{Mg\,II)}}{1000\,\kms} \right]^2
\left[\frac{\lambda L_{\lambda}(3000 \mbox{\AA})}{10^{44}\,\ergs}
\right]^{0.5}.
\end{equation}

The FWHMs of the three lines and the three continuum fluxes used above
are listed in Table~\ref{tb:fwhm}.  We adopt a combination of
measurements, using the median of the three estimates as the adopted
black hole mass for future use.  When we only have two measurements,
we use a linear average.  We provide the individual estimates and the
combined estimate in Table~\ref{tb:mbh}.  

\begin{figure}
\epsscale{0.5}
\plotone{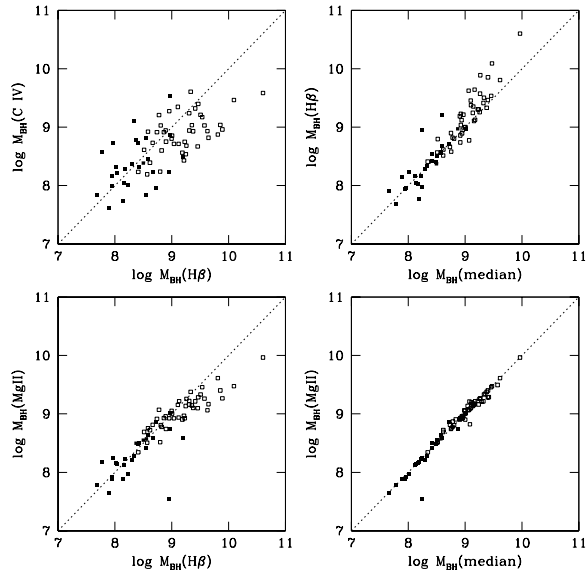}
\caption{
Comparison of black hole masses estimated using different scaling
relationships.  M$_{BH}$(median) is our adopted value.
\label{fg:mbh}}
\end{figure}

We compared the black hole masses calculating using different scaling
relationships in Figure~\ref{fg:mbh}.  We found that the estimates
from \hb\ and \mgii\ agree better than those from \civ\ and our adopted
black hole masses are mostly from the estimates using \hb\ or \mgii.

\begin{figure}
\epsscale{0.3}
\includegraphics[angle=270,scale=0.3]{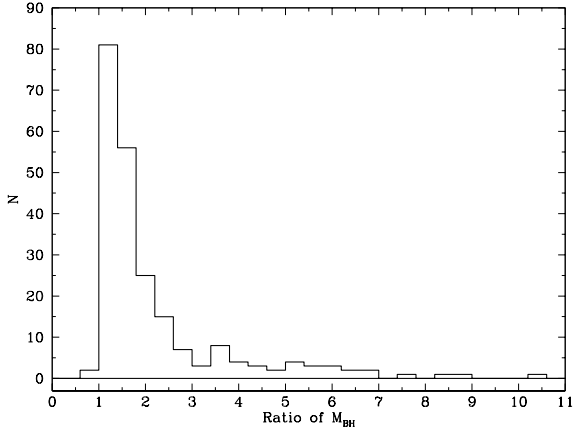}
\caption{
Distribution of the ratio of black hole masses estimated using
different scaling relationships.
\label{fg:mbhratio}}
\end{figure}

Moreover, we take the ratio of any two estimates for each object and
plot the distribution of all the ratios in Figure~\ref{fg:mbhratio}.
We note that most estimates agree within a factor of 2, but there are
1/3 of the ratios greater than 2, indicating possible larger
uncertainties of estimated black hole masses using a single scaling
relationship.  A few show even large differences of a factor of 5-10,
mostly involving an estimate from a line with bad profile or lower
signal-to-noise ratio.  
We also note that some of the large differences are caused by 6
objects (3C~47, 3C~110, 3C~175, B2~0742+31, 3C~254, PG~1704+608) 
with \hb\ broader than 10,000\,\kms.  These result in the
largest black hole masses (using \hb) in these objects, which seem to
deviate from the strong agreement between the estimates from \hb\ and
\mgii\ for lower black hole masses (Fig.~\ref{fg:mbh}).  It is unclear
what physical processes may be involved in producing the extremely
broad \hb, if they are not a natural extension of the parameter
space, but all of these largest black hole masses are excluded by the median
selection when obtaining our adopted black hole masses.
We suggest using multiple mass scaling relationships to estimate
black hole masses and adopting a reasonable combination if possible.

We are also interested in calculating the Eddington ratio, which we
define as the ratio of $\rm L_{bol}$ to $\rm L_{Edd}$, where $\rm
L_{Edd}=1.25\times 10^{38}(M_{BH}/M_{\odot})$.  We use integrated
bolometric luminosities from \citet{Runno12a}, and for the 22 objects
without measured bolometric luminosities, we use the recommended
correction of \citet{Runno12a}:
\begin{equation} \log (L_{bol}) = 4.74 + 0.91 \log (1450L_{1450}),
\label{lbol} \end{equation}
These luminosities and derived Eddington fractions are listed in
Table~\ref{tb:mbh}.
It has been suggested that the integrated bolometric luminosity should
be multiplied by 0.75 to correct for a viewing angle bias and anisotropic
disk emission
\citep[see e.g.,][]{NemBro10, Runno12a}, but we did not apply this
correction for the above quantities.
%

\section{Emission-Line Properties}
\label{sec:correlations}

We point out consistencies between our data set and some well known
correlations to highlight the fact that our sample, while
heterogeneous, does appear representative.  Still, it is likely more
appropriate for particular applications to examine particular subsamples of
Shang et al. (2011) than the entire sample.

\begin{figure}
\epsscale{0.5}
\plotone{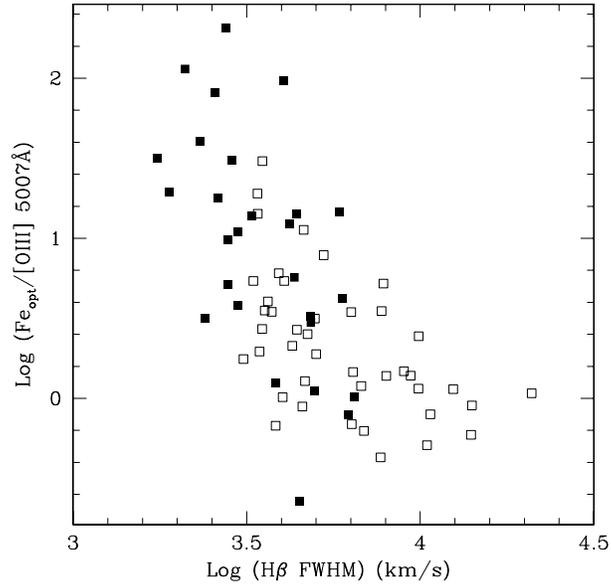}
\caption{Eigenvector 1 correlation in our sample.  Open squares are
radio-loud objects and filled squares are radio-quiet objects.
\label{fg:ev1}}
\end{figure}

The Boroson \& Green (1992) Eigenvector 1 (BGEV1) correlations are
well-studied and easily observable in many quasar samples
\citep{Wills99, Sulen00, Shang03, Zamfi10, Kovac10}.  This is the case
in our sample here, as seen in Figure~\ref{fg:ev1}, plotting the ratio
of optical \feii\ to \oiii\ against the FWHM of H$\beta$.

\begin{figure}
\epsscale{0.8}
\plotone{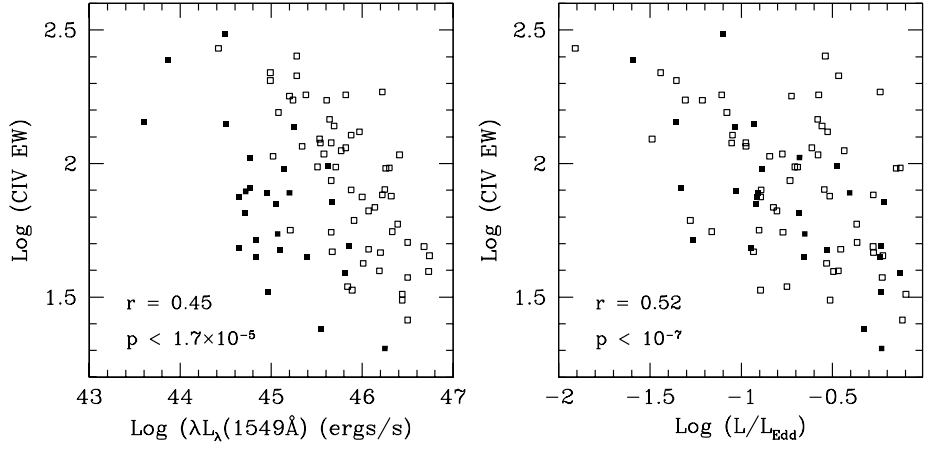}
\caption{Baldwin effect and  $\rm EW\!-\!L/L_{Edd}$ relationship in
our sample.  Open squares are
radio-loud objects and filled squares are radio-quiet objects.
The $r$ and $p$ are the 
Pearson correlation coefficient and the
two-tailed probability of a correlation arising by chance,
respectively.
\label{fg:baldwin}}
\end{figure}

We also plot the Baldwin effect for \civ\ as well as the EW of \civ\
versus the Eddington fraction (Fig.~\ref{fg:baldwin}).  Both
correlations are present in our sample, and as others have reported
the correlation is stronger in the case of the Eddington fraction
\citep{BasLao04}.

More could be done with our sample, particularly for subsamples,
although little would likely be new or ground breaking in the realm of
line-line or line-continuum correlations.  The real utility will be in
future studies in which emission-line properties are compared to SEDs,
as well as for other projects beyond the scope of the present work.
In particular, the radio-loud subsample was originally selected to
study orientation effects, which appear to impact black hole mass estimates
(Runnoe et al. 2012b).

\section{Summary}
\label{sec:conclusion}

In this paper we have reported the measurements of the emission lines 
of a sample of 85 quasars for redshifts $0<z<1.5$ with detailed
radio-to-X-ray SEDs.  The sample, although heterogeneous, appears to be
representative of bright quasars in the low-to-moderate redshift
universe for both radio-loud and radio-quiet subclasses.

We have used self-consistent scaling relationships and well-determined
bolometric luminosity corrections to estimate black hole masses and Eddington
fractions for future applications.  We have noticed that black hole
masses estimated using different scaling relationships usually agree
within a factor of 2, but a significant part (1/3) also shows larger
difference of a factor of 2-10.  We suggest that a reasonable
combination (e.g., median or average) of multiple estimates for single
objects should be pursued if possible.

The line measurements as well as the derived properties of black hole
mass and Eddington fraction will be useful for future studies
involving the detailed SEDs of the sample objects.

\acknowledgments

This work has been supported by the National Natural Science
Foundation of China (Grant No. 10773006, 11133001, and 10878010) and
Chinese 973 Program 2007CB815405.  We are also grateful for support by
Tianjin Distinguished Professor Funds and the Ph. D. Programs
Foundation of Ministry of Education of China (20100091110009).
This work was also supported by NASA through grant
HST-GO-10717.01-A, Spitzer-GO-20084, and Grant No. NNG05GD03G.



\clearpage

\begin{figure}
\epsscale{0.5}
\includegraphics[angle=270,scale=0.3]{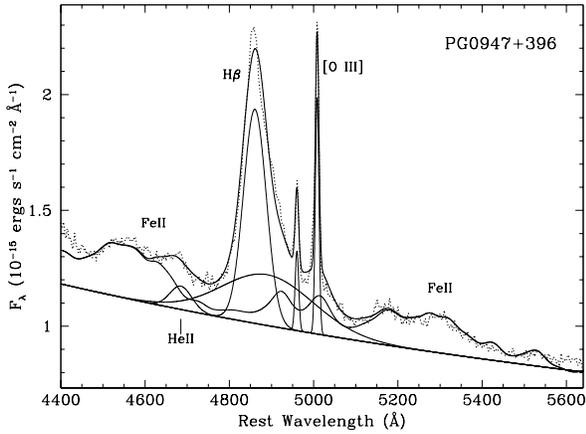}
\caption{Example of model fitting to the spectra in \hb\
region.  Shown with the data (dotted-line) are the fitting results
(thick solid line), and individual components (thin solid lines)
including two Gaussians for \hb, single Gaussian for \oiii, \feii\
template, and power-law continuum.  \heii\ is also modeled in this
object, but it is not important in most objects.
\label{fg:fitline}}
\end{figure}

\clearpage

\begin{deluxetable}{rlllllcrc}
\tabletypesize{\scriptsize}
\tablecolumns{9}
\tablewidth{0pc}
\tablecaption{The Sample \label{tb:sample}}
\tablehead{
\colhead{ID} &
\colhead{Name} & \colhead{Other Name}  & \colhead{RA(J2000)}&
\colhead{DEC(J2000)} &
\colhead{$z$\tablenotemark{a}}
& \colhead{E(B$-$V)\tablenotemark{b}}
& \colhead{R\tablenotemark{c}}
& \colhead{SampleID}
}
\startdata

 1& MC2 0042+101&  
	&00:44:58.72 & $+$10:26:53.7 & 0.5870 & 0.068
	&   840\phm{.13} & RLQ\\
 2& PG 0052+251&  
	&00:54:52.10 & $+$25:25:38.0 & 0.1544 & 0.047
	&  0.34 & FUSE\\
 3& PKS 0112$-$017& UM 310
	&01:15:17.10 & $-$01:27:04.6 & 1.3743 & 0.062
	&  2819\phm{.13} & RLQ\\
 4& 3C 37&  
	&01:18:18.49 & $+$02:58:06.0 & 0.6661 & 0.039
	&  5550\phm{.13} & RLQ\\
 5& 3C 47&  
	&01:36:24.40 & $+$20:57:27.0 & 0.4250 & 0.061
	&  6570\phm{.13} & RLQ\\
 6& 4C 01.04& PHL 1093
	&01:39:57.25 & $+$01:31:46.2 & 0.2634 & 0.029
	&  2556\phm{.13} & RLQ\\
 7& 4C 10.06& PKS 0214+10
	&02:17:07.66 & $+$11:04:10.1 & 0.4075 & 0.109
	&   472\phm{.13} & RLQ\\
 8& PKS 0403$-$13&  
	&04:05:34.00 & $-$13:08:13.7 & 0.5700 & 0.058
	&  5413\phm{.13} & RLQ\\
\enddata
\tablenotetext{a}{Redshift from \citet{Shang11}.}
\tablenotetext{b}{Galactic Extinction from NED \url{(http://nedwww.ipac.caltech.edu/)}
based on \citet{Schle98}.}
\tablenotetext{c}{Radio loudness, $R = f(5\,GHz)/f(4215\mbox{\AA})$,
calculated using our data \citep{Shang11}.}
\end{deluxetable}

\clearpage

\begin{deluxetable}{lllllrrp{0pt}@{}llp{0pt}@{}ll}
\setlength{\tabcolsep}{0.06in}
\tabletypesize{\tiny}
\tablecolumns{16}
\tablewidth{0pc}
\tablecaption{Parameters of Strong Emission Lines \lya\ and \nv. 
\label{tb:stronglinesA}}
\tablehead{
\colhead{} & 
\colhead{} & 
\multicolumn{5}{c}{\lyal} & \colhead{} &
\multicolumn{2}{c}{\nvl\tablenotemark{a}} & \colhead{} &
\\ \cline{3-7} \cline{9-10} \cline{12-13}
\colhead{ID} &
\colhead{Object} &
\colhead{Flux}&\colhead{EW}&\colhead{FWHM}&\colhead{\dv}&\colhead{Asymm} &
\colhead{} &
\colhead{Flux}&\colhead{EW}&
\colhead{} &
}
\startdata

1& MC2 0042+101 	&\phn      88$^{+7    }_{-6    }$
	&   290.1$^{+85.8}_{-61.5}$
	&   3185$^{+55     }_{-55     }$
	&$   -135$
	&$ -0.317^{+0.025  }_{-0.023  }$
&
	&\phn      1.
	&\phn     1.5
\\
2& PG 0052+251 	&    3078$^{+124  }_{-107  }$
	&   130.6$^{+11.0}_{-10.2}$
	&   3520$^{+30     }_{-30     }$
	&$   -230$
	&$ -0.065^{+0.006  }_{-0.006  }$
&
	&     126$^{+151  }_{-73   }$
	&\phn     5.5$^{+7.1 }_{-6.5 }$
\\
3& PKS 0112$-$017 	&\phn      43$^{+4    }_{-4    }$
	&\phn    52.9$^{+8.1 }_{-7.4 }$
	&   2690$^{+50     }_{-40     }$
	&$   -410$
	&$ -0.023^{+0.015  }_{-0.011  }$
&
	&\phn      10$^{+7    }_{-4    }$
	&    12.5$^{+10.0}_{-9.1 }$
\\
4& 3C 37 	&\phn     158$^{+6    }_{-5    }$
	&   320.1$^{+50.4}_{-41.1}$
	&   2710$^{+20     }_{-15     }$
	&$    -30$
	&$ -0.197^{+0.011  }_{-0.007  }$
&
	&\phn       8$^{+9    }_{-5    }$
	&    17.6$^{+22.8}_{-18.5}$
\\
5& 3C 47 	&\phn     395$^{+39   }_{-33   }$
	&   282.8$^{+102.6}_{-69.3}$
	&   4475$^{+100    }_{-100    }$
	&$   -360$
	&$ -0.199^{+0.018  }_{-0.013  }$
&
	&\phn       5
	&\phn     3.5
\\
6& 4C 01.04 	&\phn     129$^{+14   }_{-12   }$
	&   157.4$^{+49.6}_{-36.0}$
	&   4375$^{+240    }_{-225    }$
	&$   -325$
	&$ -0.195^{+0.018  }_{-0.019  }$
&
	&\phn      40$^{+22   }_{-14   }$
	&    49.6$^{+40.3}_{-29.8}$
\\
7& 4C 10.06 	&\phn     934$^{+80   }_{-67   }$
	&   133.9$^{+27.7}_{-22.7}$
	&   2670$^{+40     }_{-45     }$
	&$     75$
	&$ -0.116^{+0.017  }_{-0.016  }$
&
	&\phn      76$^{+93   }_{-44   }$
	&    11.1$^{+16.3}_{-13.3}$
\\
8& PKS 0403$-$13 	&\phn     309$^{+14   }_{-12   }$
	&   139.5$^{+15.6}_{-13.9}$
	&   3405$^{+20     }_{-25     }$
	&$    415$
	&$ -0.071^{+0.004  }_{-0.005  }$
&
	&\phn      41$^{+21   }_{-14   }$
	&    18.7$^{+11.4}_{-10.2}$
\\
\enddata
\tablecomments{Flux --- observed-frame flux in 10$^{-15}$\flux.
EW --- rest-frame equivalent width.
FWHM --- in \kms.
Asymm --- asymmetry parameter defined as 
$\mbox{Asymm} =
[{\lambda_c(\frac{3}{4}) - \lambda_c(\frac{1}{4})}] / {\mbox{FWHM}}
$ .
If the S/N ratio is too low, no reasonable flux and EW errors
can be measured, therefore the flux and EW without errors are less
reliable.
\dv\ --- line peak velocity shift (\kms) relative to the systematic redshift
. The error for \dv\ depends on the uncertainty of the
redshift .
}
\tablenotetext{a}{The values are for the sum of the doublet.  Each
single line is assumed to have the same shape as \lya\
.}
\end{deluxetable}

\clearpage

\begin{deluxetable}{lllllrrp{0pt}@{}lllrp{0pt}@{}ll}
\setlength{\tabcolsep}{0.03in}
\tabletypesize{\scriptsize}
\tablecolumns{17}
\tablewidth{0pc}
\tablecaption{Parameters of Strong Emission Lines 
\civ, \ciii, and \siiii
\label{tb:stronglinesB}}
\tablehead{
\colhead{} & 
\colhead{} & 
\multicolumn{5}{c}{\civl\tablenotemark{a}} & \colhead{} &
\multicolumn{4}{c}{\ciiil} & \colhead{} &
\multicolumn{2}{c}{\siiiil\tablenotemark{b}}
\\ \cline{3-7} \cline{9-12} \cline{14-15}
\colhead{ID} &
\colhead{Object} &
\colhead{Flux}&\colhead{EW}&\colhead{FWHM}&\colhead{\dv}&\colhead{Asymm} &
\colhead{} &
\colhead{Flux}&\colhead{EW}&\colhead{FWHM}&\colhead{\dv}&
\colhead{} &
\colhead{Flux}&\colhead{EW}
}
\startdata

1& MC2 0042+101 	&\phn      62$^{+6    }_{-5    }$
	&   218.9$^{+42.5}_{-32.0}$
	&   4195$^{+245    }_{-210    }$
	&$   -185$
	&$ -0.150^{+0.038  }_{-0.034  }$
&
	&\phn \phn       5$^{+2    }_{-2    }$
	&    21.8$^{+11.9}_{-10.1}$
	&   2760$^{+275    }_{-240    }$
	&$   -115$
&
	&\phn \phn       2$^{+2    }_{-1    }$
	&\phn     7.7$^{+10.1}_{-8.6 }$
\\
2& PG 0052+251 	&    2181$^{+122  }_{-101  }$
	&   137.6$^{+11.3}_{-10.3}$
	&   5815$^{+175    }_{-170    }$
	&$   -255$
	&$ -0.133^{+0.013  }_{-0.010  }$
&
	&\phn     261$^{+22   }_{-18   }$
	&    22.5$^{+2.3 }_{-2.2 }$
	&   3820$^{+140    }_{-130    }$
	&$   -170$
&
	&\phn      42$^{+19   }_{-14   }$
	&\phn     3.5$^{+1.7 }_{-1.7 }$
\\
3& PKS 0112$-$017 	&\phn      20$^{+2    }_{-2    }$
	&\phn    30.8$^{+4.1 }_{-3.8 }$
	&   5030$^{+550    }_{-460    }$
	&$   -325$
	&$  0.214^{+0.014  }_{-0.018  }$
&
	&\phn \phn       6$^{+1    }_{-1    }$
	&    11.6$^{+2.7 }_{-2.6 }$
	&   6085$^{+380    }_{-370    }$
	&$   -760$
&
	&\phn \phn     0.3
	&\phn     0.5
\\
4& 3C 37 	&\phn      97$^{+4    }_{-3    }$
	&   252.6$^{+21.5}_{-18.4}$
	&   3360$^{+55     }_{-50     }$
	&$   -170$
	&$ -0.124^{+0.008  }_{-0.005  }$
&
	&\phn \phn      17$^{+4    }_{-3    }$
	&    52.7$^{+19.8}_{-15.5}$
	&   3765$^{+235    }_{-215    }$
	&$     45$
&
	&\phn \phn     0.1
	&\phn     0.3
\\
5& 3C 47 	&\phn     212$^{+15   }_{-12   }$
	&   172.9$^{+17.9}_{-16.0}$
	&   5450$^{+150    }_{-145    }$
	&$   -435$
	&$  0.108^{+0.035  }_{-0.030  }$
&
	&\phn \phn      41$^{+7    }_{-6    }$
	&    49.5$^{+11.8}_{-10.5}$
	&   3765$^{+340    }_{-290    }$
	&$   -720$
&
	&\phn \phn     0.9
	&\phn     1.1
\\
6& 4C 01.04 	&\phn     174$^{+15   }_{-12   }$
	&   270.4$^{+53.3}_{-39.0}$
	&   6665$^{+380    }_{-350    }$
	&$      0$
	&$ -0.102^{+0.023  }_{-0.017  }$
&
	&\phn \phn      38$^{+9    }_{-7    }$
	&    63.0$^{+23.6}_{-19.4}$
	&   6470$^{+635    }_{-610    }$
	&$     45$
&
	&\nodata&\nodata
\\
\enddata
\tablecomments{Same as Table~\ref{tb:stronglinesA} for \civ, \ciii,
and \siiii\ emission lines.}
\tablenotetext{a}{Flux and EW are the sum of the \civ\ doublet.
FWHM of \civ\ is for a single component of the doublet}
\tablenotetext{b}{\siiii\ is assumed to have the same 
FWHM and \dv\ as \ciii\ .
}
\end{deluxetable}

\clearpage

\begin{deluxetable}{llllrp{0pt}@{}llllrrp{0pt}@{}lllr}
\setlength{\tabcolsep}{0.01in}
\tabletypesize{\tiny}
\tablecolumns{17}
\tablewidth{0pc}
\tablecaption{Parameters of Strong Emission Lines 
\mgii, \hb, and \oiii
\label{tb:stronglinesC}}
\tablehead{
\colhead{} & 
\colhead{} & 
\multicolumn{4}{c}{\mgii\tablenotemark{a}} & \colhead{} &
\multicolumn{5}{c}{\hb} & \colhead{} &
\multicolumn{4}{c}{\oiiil\tablenotemark{b}}
\\ \cline{3-6} \cline{8-12} \cline{14-17}
\colhead{ID} &
\colhead{Object} &
\colhead{Flux}&\colhead{EW}&\colhead{FWHM}&\colhead{\dv}&
\colhead{} &
\colhead{Flux}&\colhead{EW}&\colhead{FWHM}&\colhead{\dv}&\colhead{Asymm} &
\colhead{} &
\colhead{Flux}&\colhead{EW}&\colhead{FWHM}&\colhead{\dv}
}
\startdata

1& MC2 0042+101 	&\phn \phn      14$^{+3    }_{-2    }$
	&    79.1$^{+20.9}_{-18.6}$
	&   6275$^{+420    }_{-400    }$
	&$      0$
&
	&\phn \phn       8$^{+3    }_{-2    }$
	&   105.5$^{+62.1}_{-48.1}$
	&   8270$^{+1040   }_{-980    }$
	&$    -75$
	&$  0.000^{+0.020  }_{-0.031  }$
&
	&\phn \phn     3.9$^{+0.3  }_{-0.3  }$
	&    50.4$^{+12.6}_{-9.7 }$
	&\phn    606$^{+10     }_{-10     }$
	&$    -75$
\\
2& PG 0052+251 	&\phn     601$^{+69   }_{-58   }$
	&    89.7$^{+13.6}_{-12.8}$
	&   4510$^{+165    }_{-165    }$
	&$   -160$
&
	&\phn     276$^{+10   }_{-9    }$
	&   128.3$^{+5.9 }_{-5.8 }$
	&   6460$^{+60     }_{-70     }$
	&$    -15$
	&$  0.000^{+0.002  }_{-0.002  }$
&
	&   110.4$^{+0.9  }_{-0.8  }$
	&    53.8$^{+1.0 }_{-1.0 }$
	&\phn    847$^{+0      }_{-0      }$
	&$    -15$
\\
3& PKS 0112$-$017 	&\phn \phn       7$^{+2    }_{-1    }$
	&    30.1$^{+10.1}_{-9.3 }$
	&   4130$^{+255    }_{-250    }$
	&$   -100$
&
	&\nodata&\nodata&\nodata&\nodata&\nodata
&
	&\nodata&\nodata&\nodata&\nodata
\\
4& 3C 37 	&\phn \phn      25$^{+5    }_{-4    }$
	&   129.6$^{+41.2}_{-33.5}$
	&   4200$^{+180    }_{-175    }$
	&$    720$
&
	&\phn \phn      15$^{+2    }_{-2    }$
	&   272.9$^{+33.4}_{-33.4}$
	&   4280$^{+190    }_{-170    }$
	&$     20$
	&$  0.000^{+0.013  }_{-0.014  }$
&
	&\phn \phn     5.3$^{+0.2  }_{-0.2  }$
	&   105.9$^{+4.0 }_{-4.0 }$
	&\phn    904$^{+10     }_{-10     }$
	&$     20$
\\
5& 3C 47 	&\phn \phn      56$^{+9    }_{-7    }$
	&   124.1$^{+26.6}_{-24.4}$
	&   7745$^{+625    }_{-595    }$
	&$     80$
&
	&\phn \phn      26$^{+4    }_{-3    }$
	&   153.2$^{+35.6}_{-31.6}$
	&  14005$^{+755    }_{-755    }$
	&$     10$
	&$  0.000^{+0.016  }_{-0.014  }$
&
	&\phn    19.1$^{+0.3  }_{-0.3  }$
	&   116.2$^{+9.7 }_{-8.6 }$
	&   1128$^{+10     }_{-0      }$
	&$     10$
\\
\enddata
\tablecomments{Same as Table~\ref{tb:stronglinesA} for \mgii, \hb, and
\oiii\ emission
lines.}
\tablenotetext{a}{Flux and EW are the sum of the \mgii\ doublet.
FWHM of \mgii\ is for a single component of the doublet
}
\tablenotetext{b}{Non-zero \dv\ reflects the uncertainty of the
redshift \citep{Shang11}. 
}
\end{deluxetable}

\clearpage
\begin{deluxetable}{llllp{0pt}@{}llp{0pt}@{}rrp{0pt}@{}rr}
\tabletypesize{\tiny}
\tablecolumns{17}
\tablewidth{0pc}
\tablecaption{Parameters of Strong Emission Lines \lya, \nv, and}
\tablecaption{Parameters of
UV and Optical \fe\ Blends, and the Small Blue Bump
\label{tb:stronglinesD}}
\tablehead{
\colhead{} & 
\colhead{} & 
\colhead{} & 
\colhead{} & \colhead{} &
\colhead{} &
\colhead{} &
\colhead{} &
\multicolumn{5}{c}{Small Blue Bump\tablenotemark{c}}
\\ \cline{9-13}
\colhead{} & 
\colhead{} & 
\multicolumn{2}{c}{Optical \feii\tablenotemark{a}} &
\colhead{} &
\multicolumn{2}{c}{UV Fe\tablenotemark{b}} & 
\colhead{} &
\multicolumn{2}{c}{local cont.} & 
\colhead{} &
\multicolumn{2}{c}{global cont.}
\\ \cline{3-4} \cline{6-7} \cline{9-10} \cline{12-13}
\colhead{ID} &
\colhead{Object} &
\colhead{Flux}&\colhead{EW}& 
\colhead{} &
\colhead{Flux}&\colhead{EW}& 
\colhead{} &
\colhead{Flux}&\colhead{EW}& 
\colhead{} &
\colhead{Flux}&\colhead{EW} 
}
\startdata

1& MC2 0042+101 	&\nodata&\nodata
&
	&\phn      20$^{+9    }_{-9    }$
	&\phn    40.0$^{+61.1}_{-55.3}$
&
	&      51&   345&
	&      88&   689
\\
2& PG 0052+251 	&\phn     113$^{+14   }_{-14   }$
	&\phn    52.6$^{+7.8 }_{-7.4 }$
&
	&    1144$^{+249  }_{-249  }$
	&\phn    45.0$^{+44.0}_{-41.3}$
&
	&    2475&   476&
	&    3727&   817
\\
3& PKS 0112$-$017 	&\nodata&\nodata
&
	&\phn      17$^{+5    }_{-5    }$
	&\phn    83.0$^{+23.9}_{-22.5}$
&
	&\nodata&\nodata&
	&      97&   421
\\
4& 3C 37 	&\phn \phn      11$^{+3    }_{-3    }$
	&   204.1$^{+81.5}_{-70.0}$
&
	&\phn      43$^{+15   }_{-15   }$
	&\phn    13.0$^{+106.5}_{-90.2}$
&
	&      69&   442&
	&      98&   690
\\
5& 3C 47 	&\phn \phn      11$^{+4    }_{-4    }$
	&\phn    67.1$^{+30.5}_{-27.0}$
&
	&\phn      60$^{+19   }_{-19   }$
	&\phn    15.0$^{+49.6}_{-45.6}$
&
	&     125&   339&
	&      79&   200
\\
6& 4C 01.04 	&\phn \phn      34$^{+11   }_{-11   }$
	&\phn    95.0$^{+48.4}_{-38.2}$
&
	&\phn      57$^{+36   }_{-36   }$
	&\phn    16.0$^{+80.1}_{-70.5}$
&
	&     208&   516&
	&     171&   409
\\
7& 4C 10.06 	&\phn \phn      45$^{+6    }_{-6    }$
	&\phn    59.9$^{+8.3 }_{-8.2 }$
&
	&\phn     279$^{+182  }_{-182  }$
	&\phn    18.0$^{+96.0}_{-83.9}$
&
	&     763&   443&
	&    1013&   640
\\
8& PKS 0403$-$13 	&\phn \phn      23$^{+7    }_{-7    }$
	&\phn    67.1$^{+19.4}_{-19.5}$
&
	&\phn      90$^{+36   }_{-36   }$
	&\phn    84.0$^{+56.8}_{-50.6}$
&
	&     160&   249&
	&     111&   165
\\
\enddata
\tablecomments{Same as Table~\ref{tb:stronglinesA} for UV and optical
Fe emissions, and the Small Blue Bump.}
\tablenotetext{a}{Flux is from the template fitting between 4478--5640\AA.
EW is calculated using the local continuum at 4861\AA.}
\tablenotetext{b}{Flux is from the template fitting between 2230--3016\AA.
EW is calculated using the local continuum at 2799\AA.}
\tablenotetext{c}{The small blue bump is not discussed in this paper. See Shang et al.~(2007) for descriptions.}

\end{deluxetable}

\clearpage

\begin{deluxetable}{lllllrp{0pt}@{}lllrr}
\tabletypesize{\scriptsize}
\setlength{\tabcolsep}{0.05in}
\tablecolumns{12}
\tablewidth{0pc}
\tablecaption{Parameters of Strong Emission Lines 
\nai+\hei\ and \ha
\label{tb:stronglinesE}}
\tablehead{
\colhead{} & 
\colhead{} & 
\multicolumn{4}{c}{\nail\ + \heil} & \colhead{} &
\multicolumn{5}{c}{\ha} 
\\ \cline{3-6} \cline{8-12} 
\colhead{ID} &
\colhead{Object} &
\colhead{Flux}&\colhead{EW}&\colhead{FWHM}&\colhead{\dv\tablenotemark{a}}&
\colhead{} &
\colhead{Flux}&\colhead{EW}&\colhead{FWHM}&\colhead{\dv}&\colhead{Asymm} 
}
\startdata

1& MC2 0042+101 	&\nodata&\nodata&\nodata&\nodata
&
	&\nodata&\nodata&\nodata&\nodata&\nodata
\\
2& PG 0052+251 	&\phn    49.3$^{+9.0  }_{-7.3  }$
	&    28.5$^{+5.6 }_{-5.5 }$
	&   4745$^{+345    }_{-300    }$
	&$   -100$
&
	&\phn   759.3$^{+6.9  }_{-6.1  }$
	&   446.5$^{+9.4 }_{-9.2 }$
	&   4465$^{+10     }_{-10     }$
	&$    -70$
	&$  0.000^{+0.001  }_{-0.001  }$
\\
3& PKS 0112$-$017 	&\nodata&\nodata&\nodata&\nodata
&
	&\nodata&\nodata&\nodata&\nodata&\nodata
\\
4& 3C 37 	&\nodata&\nodata&\nodata&\nodata
&
	&\nodata&\nodata&\nodata&\nodata&\nodata
\\
5& 3C 47 	&\phn \phn     2.0$^{+5.4  }_{-3.6  }$
	&    12.6$^{+40.4}_{-31.4}$
	&   6815$^{+1175   }_{-0      }$
	&$     45$
&
	&\phn \phn    98.4$^{+9.4  }_{-7.4  }$
	&   612.2$^{+153.7}_{-119.7}$
	&   5125$^{+295    }_{-255    }$
	&$   -100$
	&$  0.000^{+0.101  }_{-0.112  }$
\\
6& 4C 01.04 	&\nodata&\nodata&\nodata&\nodata
&
	&\phn   120.0$^{+9.4  }_{-7.4  }$
	&   301.4$^{+40.8}_{-36.9}$
	&   8335$^{+130    }_{-130    }$
	&$   2075$
	&$  0.000^{+0.013  }_{-0.012  }$
\\
7& 4C 10.06 	&\phn \phn     9.0$^{+5.5  }_{-3.7  }$
	&    13.7$^{+9.2 }_{-8.6 }$
	&   4560$^{+1050   }_{-810    }$
	&$    190$
&
	&\phn   335.6$^{+21.1 }_{-17.6 }$
	&   507.8$^{+66.6}_{-59.0}$
	&   4720$^{+60     }_{-55     }$
	&$    260$
	&$ -0.056^{+0.009  }_{-0.011  }$
\\
8& PKS 0403$-$13 	&\nodata&\nodata&\nodata&\nodata
&
	&\phn \phn    86.1$^{+1.0  }_{-0.9  }$
	&   316.9$^{+3.6 }_{-3.6 }$
	&   3320$^{+5      }_{-10     }$
	&$    220$
	&$ -0.139^{+0.002  }_{-0.001  }$
\\
\enddata
\tablecomments{Same as Table~\ref{tb:stronglinesA} for \nai+\hei\ and \ha\
emission lines.}
\tablenotetext{a}{Calculated using \heil\ wavelength.}
\end{deluxetable}

\clearpage

\begin{deluxetable}{llcccrrr}
\tabletypesize{\scriptsize}
\tablecolumns{8}
\tablewidth{0pc}
\tablecaption{Measured FWHM and Flux Density
\label{tb:fwhm}
}
\tablehead{
\colhead{} &
\colhead{} &
\multicolumn{3}{c}{FWHM (\kms)} &
\colhead{} &
\colhead{} &
\colhead{}
\\\cline{3-5}
\colhead{ID} &
\colhead{Object} &
\colhead{\mgii} &
\colhead{\hb} &
\colhead{\civ} &
\colhead{$f_{1350}$\tablenotemark{a}} &
\colhead{$f_{3000}$\tablenotemark{a}} &
\colhead{$f_{5100}$\tablenotemark{a}}
}
\startdata

1& MC2 0042+101
	& 6275 & 8270 & 4195
	&  0.32 &  0.16 &  0.08\\
2& PG 0052+251
	& 4510 & 6460 & 5815
	& 18.96 &  6.06 &  1.99\\
3& PKS 0112$-$017
	& 4130 &\nodata & 5030
	&  0.75 &  0.21 &  0.09\\
4& 3C 37
	& 4200 & 4280 & 3360
	&  0.44 &  0.18 &  0.05\\
5& 3C 47
	& 7745 &14005 & 5450
	&  1.57 &  0.41 &  0.16\\
6& 4C 01.04
	& 9550 & 9905 & 6665
	&  0.74 &  0.53 &  0.35\\
7& 4C 10.06
	& 4670 & 4735 & 3785
	&  6.60 &  2.04 &  0.70\\
8& PKS 0403$-$13
	& 3535 & 3735 & 3325
	&  1.77 &  0.71 &  0.32\\
\enddata
\tablenotetext{a}{Observed-frame flux density at rest-frame 1350, 3000, and
5100\,\AA\ in the units of $10^{-15}$\fluxl, from fitted local
continuum in each region.}
\end{deluxetable}

\clearpage

\begin{deluxetable}{llrrrcccc}
\tabletypesize{\scriptsize}
\tablecolumns{9}
\tablewidth{0pc}
\tablecaption{Luminosity, Black Hole Mass, and Eddington Ratio
\label{tb:mbh}
}
\tablehead{
\colhead{} &
\colhead{} &
\multicolumn{4}{c}{$\log \mbh$ ($\msun$)} &
\colhead{} &
\colhead{} &
\colhead{}
\\\cline{3-6}
\colhead{ID} &
\colhead{Object} &
\colhead{\mgii} &
\colhead{\hb} &
\colhead{\civ} &
\colhead{Adopted\tablenotemark{a}} &
\colhead{$\log$(\lbol)\tablenotemark{b}} &
\colhead{$L/L_{Edd}$} &
\colhead{\lbol\ Note\tablenotemark{c}}
}
\startdata

1& MC2 0042+101
	& 8.97 & 9.21 & 8.43 & 8.97
	&45.68 & 0.04 &calc\\
2& PG 0052+251
	& 8.74 & 8.97 & 8.87 & 8.87
	&45.94 & 0.09 &\\
3& PKS 0112$-$017
	& 9.21 &\nodata & 9.36 & 9.29
	&46.88 & 0.31 &\\
4& 3C 37
	& 8.72 & 8.62 & 8.39 & 8.62
	&46.18 & 0.29 &\\
5& 3C 47
	& 9.17 & 9.64 & 8.82 & 9.17
	&45.97 & 0.05 &\\
6& 4C 01.04
	& 9.14 & 9.25 & 8.54 & 9.14
	&45.44 & 0.02 &\\
7& 4C 10.06
	& 9.05 & 9.00 & 8.81 & 9.00
	&46.48 & 0.24 &\\
8& PKS 0403$-$13
	& 8.78 & 8.82 & 8.60 & 8.78
	&46.36 & 0.30 &\\
\enddata
\tablenotetext{a}{The adopted \mbh\ is the median of 3 available estimates
or the mean in case of only 2 estimates.}
\tablenotetext{b}{From \citet{Runno12a}. \lbol\ is in \ergs,
calculated using the full radio-to-X-ray SEDs (63 objects) 
except for the ones noted in the last column (22 objects).}
\tablenotetext{c}{Those marked with "calc" indicate that \lbol\ is
calculated using equation~\ref{lbol} because the SEDs of these 22
objects do not have adequate X-ray or mid-infrared coverage
\citep{Runno12a}. }
\end{deluxetable}

\label{lastpage}

\end{document}